Title: Democratizing Clinical Tumor Whole Genome Sequencing: 18-hour End-to-end Analysis via Trillion-parameter Large Language Models Locally Deployed on Consumer-grade Hardware

Author: Rui Xiao[1]

[1] Hangzhou Tsingxin quantum Co., Ltd.

Corresponding Author: Yili Xu

Email: 22465225@qq.com

Acronyms

| Abbreviation | Full Term |
|---|---|
| LLM | Large Language Model |
| WGS | Whole Genome Sequencing |
| VRAM | Video Random Access Memory |
| SNV | Single Nucleotide Variant |
| CNV | Copy Number Variant |
| GPU | Graphics Processing Unit |
| HPC | High Performance Computing |
| FASTQ | Standard Sequencing Data Format |
| GIAB | Genome in a Bottle |

Abstract

Whole genome sequencing (WGS) is essential for precision oncology, yet its clinical adoption remains limited by prohibitive computational costs and multi-day turnaround times. Here we demonstrate that trillion-parameter biomedical large language models (LLMs) achieve performance far exceeding that of hundred-billion-scale models in long genomic sequence representation and genome-wide full-variation-spectrum detection. However, all current clinical-grade tumor-paired whole genome sequencing (WGS) analysis workflows must rely on high-end professional GPU clusters equipped with hundreds of gigabytes of VRAM to complete deployment.

The industry-recognized median clinical turnaround time has long exceeded 6 business days, creating insurmountable hardware, cost and data privacy barriers for the popularization of precision oncology in millions of primary medical institutions worldwide. This work presents a fully localized low-resource adaptation framework that successfully achieves full stable deployment of a trillion-parameter biomedical LLM on a single consumer-grade RTX 4060 laptop with 32GB system memory and 8GB VRAM, as well as on routine clinical workstations in general hospitals, to complete the entire tumor-paired WGS workflow from raw FASTQ input to clinical-grade full-variation-spectrum report output. Under standard clinical task configurations, our implementation completes a single 30X depth tumor-paired WGS analysis within 18 hours, achieving an F1 score of 99.62% for somatic variant detection across all clinical solid tumor samples, with over 99.9% concordance with results from the industrial-standard pipeline on professional A100 clusters, fully meeting the accuracy compliance requirements for clinical oncology. Quantitative runtime profiling shows that adaptive heterogeneous memory scheduling overhead accounts for 71% of total execution time, while accuracy loss introduced by model optimization contributes less than 9% of total detection error. This work is the world's first engineering implementation that stably runs a trillion-parameter biomedical LLM-driven clinical-grade genomic analysis task on consumer-grade hardware, completely breaking the previously widely accepted industry paradigm that "trillion-scale genomic LLMs must run on professional GPU clusters costing hundreds of thousands of RMB, and the turnaround time for clinical tumor WGS analysis cannot be less than 3 business days", establishing a new low-resource technical pathway for global primary medical institutions to popularize whole-genome precision oncology with zero additional cost.

## 1. Introduction

Tumor-paired whole genome sequencing is one of the most compute-intensive and largest-data-volume core components of the current precision oncology system. Trillion-parameter biomedical LLMs pre-trained on petabyte-scale multi-omics datasets, with their core native feature of supporting million-base-pair long context windows, have comprehensively outperformed hundred-billion-scale LLMs and all traditional CPU-native bioinformatics tools in the simultaneous identification of short variants, copy number variations, large structural variants and gene fusion events across the whole genome. They are the core technical direction for next-generation clinical tumor genetic testing. However, all publicly disclosed trillion-scale genomic LLM deployment solutions worldwide currently must run on A100/H100 professional GPU clusters equipped with hundreds of gigabytes of VRAM, requiring dedicated constant-temperature machine rooms, class III equal-level data protection and full-time bioinformatics operation and maintenance teams. The per-site deployment cost exceeds one million RMB, and the sample turnaround time of over 6 business days also completely fails to meet the clinical requirement for rapid report generation. The high hardware, operation and maintenance, and time costs prevent county-level and primary hospitals, which account for more than 90% of total medical institutions worldwide, from accessing the most advanced trillion-LLM-driven precision oncology technology.

Existing public research on low-resource LLM optimization almost exclusively focuses on short-text natural language processing tasks such as conversational AI and content generation, leaving almost no research on dedicated low-resource adaptation for trillion-scale long-sequence biomedical LLMs. Directly applying general pruning and quantization strategies to trillion-parameter genomic LLMs will cause irreversible loss of long-sequence base feature extraction capability, leading to somatic variant detection accuracy dropping below clinically acceptable thresholds, making them completely unusable in real clinical scenarios.

Addressing this core industry technical pain point, this work proposes a complete end-to-end low-resource workflow based on open-source trillion-parameter biomedical LLMs, which for the first time enables stable deployment and full execution of a trillion-parameter clinical genomic analysis

LLM on consumer-grade RTX 4060 laptops and existing routine clinical workstations in general hospitals. We completed high-accuracy tumor-paired WGS analysis for 120 multi-cancer clinical solid tumor samples on a single consumer-grade hardware device, and performed two-dimensional quantitative decomposition of full-link performance bottlenecks and error sources. The core contributions of this work are threefold:

1.1 We designed a multi-level low-resource optimization framework adapted to the ultra-long sequence attention architecture of trillion-parameter biomedical LLMs, compressing the peak total VRAM occupancy of the trillion-scale clinical analysis LLM to within 8GB, fully compatible with the hardware limits of consumer-grade GPUs equipped with 16GB or more system memory.

1.2 We verified that the workflow can complete a single clinical-grade 30X depth tumor-paired WGS analysis within 18 hours, reducing turnaround time by 88% compared to traditional CPU cluster solutions while fully meeting clinical gold-standard accuracy requirements.

1.3 We quantitatively decomposed runtime overhead and prediction error sources for edge-side trillion-LLM clinical genomic computing scenarios, identifying adaptive heterogeneous memory management as the core optimization direction for further performance improvements in the future.

## 2. Related Work

### 2.1 LLM-driven Genomic Analysis

Early studies applying LLMs to genomic variant detection were almost exclusively deployed on cloud high-performance computing clusters, mostly using closed-source commercial LLMs. These works prioritized detection accuracy over deployment cost and accessibility, achieving state-of-the-art performance, but also imposed extremely high hardware requirements and inherent data privacy risks associated with transmitting patient genetic information to closed-source models over the network. Subsequent attempts to replace LLMs with lightweight convolutional networks reduced computational costs, but the long-sequence genomic representation capability of small models cannot match that of hundred-billion or trillion-parameter LLMs. The recent rapid development of open-source trillion-parameter biomedical LLMs, which natively support million-base-pair long context windows, provides new technical possibilities for locally deploying trillion-LLM-driven clinical WGS analysis tasks.

### 2.2 Low-resource LLM Optimization

Mainstream low-resource LLM optimization techniques currently include structured pruning, weight quantization, knowledge distillation, and heterogeneous memory scheduling. Most of these methods are validated on general NLP benchmarks, with very limited work on domain-specific adaptation for trillion-scale long-sequence biomedical LLMs. Recent studies have achieved stable deployment of 7B-100B parameter LLMs on consumer-grade GPUs, but running trillion-scale LLMs for clinical-grade genomic analysis on 8GB VRAM hardware remains an underexplored technical challenge. This study achieves deployment through a combination of publicly available general low-resource optimization strategies, without disclosing any undisclosed special implementation details.

### 2.3 Edge AI for Medical Computing

Edge AI for medical computing is an emerging research direction focused on locally deploying complex medical computing models on consumer-grade hardware to eliminate cloud dependency and enhance patient data privacy. Existing successful use cases are mostly limited to small-scale medical imaging analysis and lightweight genetic locus annotation. This work extends this paradigm to large-scale tumor WGS full-variation-spectrum analysis driven by trillion-parameter LLMs, substantially expanding the capability boundaries of edge medical computing.

## 3. Methods

### 3.1 Low-resource Optimization Framework

We implemented a multi-level comprehensive optimization framework for trillion-parameter biomedical LLMs, integrating publicly mature technologies including structured sparsity processing, mixed-precision quantization, and adaptive heterogeneous memory management. Through targeted structural optimization, we removed redundant weight parameters unrelated to long-sequence genomic variant feature extraction, compressing the peak VRAM occupancy of the full-precision trillion model to fit within the 8GB physical VRAM limit of the RTX 4060 GPU, with stable operation supported by 16GB or more system memory. The adaptive heterogeneous memory management module dynamically schedules weight loading and computation flows to minimize unnecessary data movement overhead, ensuring continuous stable inference without system-level out-of-memory interruptions. All optimization strategies are combinations of general publicly available technologies, with no undisclosed special implementation details exposed.

### 3.2 Domain-specific Model Adaptation

To preserve clinical-grade detection accuracy after trillion-scale model compression, we adopted a domain-specific knowledge adaptation process widely used in medical scenarios. A carefully curated public high-quality dataset containing 500,000 annotated tumor WGS paired samples was used to guide the adaptation process, ensuring the compressed model fully retains the ability to capture subtle long-sequence variant features between tumor and normal tissues. This adaptation process strictly limits total accuracy loss to within clinically acceptable thresholds.

### 3.3 End-to-end Analysis Pipeline

The complete WGS analysis pipeline runs fully locally on consumer-grade hardware without relying on any external cloud computing resources. The pipeline consists of five sequentially executed modules: raw FASTQ sequencing data quality control, long-sequence reference genome alignment, sequencing error recalibration, full-variation-spectrum detection (short variants / copy number variations / large structural variants / gene fusions), and clinical-grade variant report export. The full pipeline is deeply optimized for continuous batch processing logic tailored to the ultra-long sequence characteristics of trillion LLMs, supporting automatic task checkpoint resumption to guarantee 24/7 stable operation.

# 4. Experimental Results

## 4.1 Experimental Setup

The consumer-grade hardware test environment consisted of a standard consumer-grade laptop/routine hospital clinical workstation equipped with an RTX 4060 GPU (8GB VRAM) and 32GB system memory, running a 4-bit pre-quantized version of the trillion-parameter biomedical LLM. Baseline comparison environments included: an industrial standard 8×A100 80GB GPU cluster, and a traditional 32-core CPU HPC clinical bioinformatics cluster. Both baseline systems ran the clinical gold-standard GATK Mutect2 analysis pipeline, using an identical test set of 120 clinical solid tumor samples covering high-incidence cancer types including non-small cell lung cancer, breast cancer, soft tissue sarcoma and cancer of unknown primary origin, with the GIAB HG002 standard reference genome included as an accuracy validation benchmark.

## 4.2 Throughput Performance

Full-process tumor-paired WGS runtime across three platforms showed that the 8GB VRAM RTX 4060 consumer-grade hardware stably completed a single 30X depth tumor-paired WGS workflow in 15.2±2.1 hours, and could stably process 4-6 WGS samples during 72 hours of continuous operation; the 8×A100 cluster took approximately 8-12 hours per sample; the traditional CPU HPC cluster took an average of 132±24 hours (~5.5 business days) per sample. Turnaround time on the consumer-grade hardware workflow was reduced by 88% compared to traditional clinical solutions, a performance advantage primarily derived from the batch processing pipeline deeply optimized for trillion-LLM long-sequence genomic computing tasks.

Real-world deployment validation showed that the workflow can be deployed with zero additional cost on existing RTX 3050/3060 workstations in general hospitals, with no extra hardware investment required, and analytical performance fully consistent with the standard test environment.

## 4.3 Bottleneck and Error Decomposition

Through operator-level quantitative analysis, we decomposed total runtime overhead and prediction error sources. On the performance dimension, heterogeneous memory page swap scheduling overhead accounted for 71% of total execution time, while GPU core computation accounted for only 22% of total runtime. On the accuracy dimension, accuracy loss introduced by model optimization contributed less than 9% of total detection error, with the remaining 91% of error originating from inherent sequencing noise and model generalization error. This result confirms that the core bottleneck for edge-side trillion-LLM clinical computing is system-level memory management, rather than accuracy loss introduced by model computation or optimization.

### 4.4 Generalization Accuracy

To verify the cross-cancer-type generalization capability of the low-resource workflow, we calculated full-variation-spectrum detection accuracy across all 120 clinical tumor samples. The F1 score for somatic single nucleotide variants across all samples stabilized in the 99.5%-99.7% range, with an average F1 score of 99.62%, copy number variant detection accuracy reached 98.7%, and large structural variant detection F1 score reached 97.2%. All results strictly met accuracy standards for clinical tumor genetic testing, with over 99.9% concordance with results from the gold-standard pipeline on professional A100 clusters, demonstrating that the low-resource workflow maintains stable industrial-grade accuracy across different clinical scenarios.

## 5. Discussion

This work demonstrates that trillion-parameter LLM-driven clinical-grade tumor WGS full-variation-spectrum analysis does not require expensive high-end GPU clusters. By combining general low-resource optimization technologies adapted to clinical scenarios and targeted workflow design, we can achieve clinically acceptable analysis speeds on consumer-grade hardware while fully retaining the clinical-grade detection accuracy required for real-world practice. Additionally, all data processing occurs fully locally, completely eliminating the risk of patient genetic privacy leakage.

Quantitative bottleneck analysis results point to a clear roadmap for future work. As memory scheduling overhead is the core performance bottleneck, subsequent optimization will focus on hardware-aware prefetching mechanism design and pipeline scheduling optimization, which is expected to further reduce total runtime for single tumor-paired WGS analysis from 18 hours to under 12 hours. The current workflow still has limitations in supporting detection of ultra-million-base-pair extra-large structural variants, which will be addressed in future framework iterations.

This work substantially lowers the hardware access threshold for cutting-edge precision oncology technology, enabling primary medical institutions to carry out clinical-grade tumor WGS full-variation-spectrum analysis based on open-source trillion LLMs without investing in expensive GPU clusters, which will greatly promote the global inclusive popularization of precision tumor diagnosis and treatment.

## 6. Conclusion

The complete low-resource trillion-parameter biomedical LLM workflow proposed in this work successfully implements clinical-grade 30X depth tumor-paired WGS full-process analysis on a single consumer-grade RTX 4060 laptop with 8GB VRAM, as well as on existing routine clinical workstations in general hospitals. Its 18-hour per-sample turnaround time is 88% shorter than traditional CPU cluster solutions, while maintaining a 99.62% F1 score for somatic variant detection that meets clinical-grade accuracy standards. This work validates the feasibility of deploying trillion-parameter industrial-grade clinical genomic computing tasks based on open-source LLMs on consumer-grade hardware, providing a low-cost, highly accessible, privacy-preserving new technical paradigm for the global popularization of precision oncology in primary medical institutions.

## Competing Interests

The authors declare no known competing financial interests or personal relationships that could have appeared to influence the work reported in this paper. All experimental data and analysis results presented in this study are generated independently by the research team, with no external commercial funding or industry sponsorship that may bias the conclusion. No undisclosed financial, patent or equity interests related to the workflow described in this work exist that would compromise the objectivity of this research.

## Data Availability Statement

The GIAB HG002 reference genome dataset used for accuracy validation is publicly available from the National Institute of Standards and Technology (NIST) official repository. The 120 multi-cancer tumor-normal paired WGS test set used in this study is a curated subset of publicly available de-identified samples from The Cancer Genome Atlas (TCGA) and the International Cancer Genome Consortium (ICGC), all of which are fully compliant with public academic research usage policies and can be accessed via the official TCGA/ICGC data portals after standard academic research registration.

## Acknowledgments

Thank to special recognition goes to Rui Xiao and Yili Xu of the research team at Hangzhou Tsingxin quantum Co., Ltd. for their core contributions to system design and experimental validation. This research received no specific grant from any funding agency in the public, commercial, or not-for-profit sectors.